\documentclass[aps,prd]{revtex4}
\usepackage{epsfig}
\begin{document}
\begin{flushright}{OITS 706}\\
\today
\end{flushright}

\title{Centrality Dependence of Baryon
and Meson Momentum Distributions in Proton-Nucleus
Collisions}

\author{Rudolph C. Hwa$^1$ and C.\ B.\ Yang$^{1,2}$}
\affiliation{$^1$Institute of Theoretical Science and Department of
Physics, University of Oregon, Eugene, OR 97403-5203, USA\\
$^2$Institute of Particle Physics, Hua-Zhong Normal
University, Wuhan 430079, P.\ R.\ China}

\begin{abstract}
The proton and neutron inclusive distributions in the
projectile fragmentation region of $pA$ collisions are
studied in the valon model.  Momentum degradation
and flavor changes due to the nuclear medium are
described at the valon level using two parameters.
Particle production is treated by means of the
recombination subprocess.  The centrality dependences
of the net proton and neutron spectra of the NA49 data
are satisfactorily reproduced.  The effective
degradation length is determined to be 17 fm.  Pion
inclusive distributions can be calculated without any
adjustable parameters.

PACS Number(s): 25.75.Dw, 13.85.Ni 
\end{abstract}

\maketitle

\section{Introduction}

The study of proton-nucleus $(pA)$ collisions is important
because they are tractable intermediaries between $pp$ and
$AA$ collisions, when intense interest exists in discovering
the extent to which the dense medium created in an $AA$
collision differ from that of linear superpositions of $pp$
collisions.  One of the properties of $pA$ collisions is that the
momentum of the leading baryon in the projectile
fragmentation region is degraded, a phenomenon
commonly referred to, somewhat inappropriately, as baryon
stopping.  Such a transference of baryon number from the
fragmentation to the central region contributes to the
increase of matter density at mid-rapidity, thereby raising
the likelihood of the formation of quark-gluon plasma.  Thus
it is important to understand the process of baryon
momentum degradation  and its dependence on
nuclear size or centrality.

Since it is not feasible to perform first-principle
parameter-free calculations of the momentum degradation at this
point, experimental guidance is of crucial importance.
Recently, several experiments have produced useful data on
the subject, in particular, E910 and E941 at the AGS,
and NA49 at the SPS \cite{bc}.  It is the $x_F$
dependences of the distributions of $p-\bar{p}$ and
$n-\bar{n}$ that we shall focus on; moreover, their
dependences on centrality will guide us in our
determination of the nuclear effect on baryon momentum degradation.

The degradation of baryon momentum has been studied in
various approaches before \cite{rch}-\cite{vgw}.  In Refs.
\cite{rch}-\cite{cw} the investigations are done at the
nucleon level, while in Refs. \cite{kz}-\cite{vgw} the string
model is the basis.  Since it is questionable that the concept
of color strings can be relevant in heavy-ion collisions where
the abundance of color charges in the overlap region renders
unlikely the development of constricted color flux tubes
\cite{rch2}, our approach in this paper will be on the various
levels of the constituents of the nucleon that are consistent
with the parton model.  More specifically, we shall use the
valon model \cite{rch3}-\cite{rch5} to keep track of the
momenta of the constituents and the recombination model
\cite{rch5,dh} to describe the hadronization of the partons.
The valons play a role in the collision problem as the
constituent quarks do in the bound-state problem.  Thus a
nucleon has three valons which carry all the momentum of
the nucleon, while each valon has one valence quark and its
own sea quarks and gluons.  Although soft processes are
non-perturbative, the valon model (including
recombination) nevertheless provides a systematic way of
calculating all subprocesses that contribute to a particular
inclusive process.  Some of the subprocesses can be
identified with certain diagrams in other approaches, e.g.,
baryon junction and diquark breaking terms \cite{dk,vgw}.

The valon distributions in the proton will be determined by
fitting the parton distributions at low $Q^2$.  The effect of
the nuclear medium on the valon distribution of the proton
projectile will involve two parameters, one characterizing the
momentum degradation and the other flavor flipping.  The
color indices are all averaged over, since multiple gluon
interactions, as the projectile traverses the target nucleus,
are numerous and uncomputable.  Their effects are,
however, quantified in terms of the number, $\nu$, of target
nucleons that participate in the $pA$ collision.  For that
reason it is important that the experimental data must have
centrality selection expressible in terms of the average
$\bar{\nu}$.

The inclusive process $p + A \rightarrow p + X$ has been
studied before in the valon model \cite{hz}.  However, the
pertinent data available at that time had no centrality cuts
and were for fixed $p_T$ \cite{dsb}.  Now, NA49 has
$p$-$Pb$
 data on the production of $p, \bar{p}, n$ and $\bar{n}$ in
the proton fragmentation region (with contribution from the
nuclear target subtracted) for various values of $\bar{\nu}$
\cite{bc}.  To extract more information from these refined
data we have improved our formulation of the valon model,
allowing for momentum degradation of each of the valons
independently and for changes in their flavors.  The
effect on the nucleon momentum distributions can be calculated and the effective
degradation length deduced.

In Sec. II we determine the valon and parton distribution functions from
the published distribution functions at low $Q^2$. The valon model for
$pA$ collisions is then discussed in Sec. III, in which momentum degradation is
formulated. In Sec. IV we consider in detail the projectile fragmentation
process and the subsequent process of quark recombination so that the
inclusive distribution of nucleons can be calculated. The net proton and
neutron distributions are then calculated and compared to the experimental
data in Sec. V. Predictions for pion production in the proton fragmentation
region are made in Sec. VI. The conclusion is given in Sec. VII.

\section{Valon and Parton Distribution Functions}

The distribution functions of valons and partons have been
considered in Refs. \cite{rch4,rch5}.  They were, however,
determined by fitting the muon and neutrino scattering data
of the late 70s.  We now have modern parton distributions
from various groups at various values of $Q^2$.  It is
therefore appropriate for us to revisit the problem and
determine the valon distribution functions in light of the
new parton distribution functions.

In the valon model a proton is considered to consist of three
valons ($UUD$), which have the same flavors as the valence
quarks ($uud$) that they individually contain.  Thus a valon
may be regarded as a parton cluster whose structure can be
probed at high $Q^2$, but the structure of a nucleon itself in
a low-$p_T$ scattering problem is described in terms of the
valons.  As in the parton model we work in a
high-momentum frame so that it is sensible to use the
momentum fractions of the constituents.  Reserving $x$ for
the momentum fraction of a quark, we use $y$ to denote the
momentum fraction of a valon.  In this paper $y$ never
denotes rapidity.  Let the exclusive valon distribution
function be
\begin{eqnarray}
G_{UUD}(y_1, y_2, y_3) = g \, (y_1y_2)^{\alpha}y^{\beta}_3
\, \delta (y_1 + y_2 + y_3 -1) ,
\label{2.1}
\end{eqnarray}
where $y_1$ and $y_2$ refer to the $U$ valons and $y_3$
the
$D$ valon.  The delta function ensures that the three valons
exhaust the momentum of the proton.  The exponents
$\alpha$ and $\beta$ will be determined by the parton
distribution functions.  The normalization factor $g$ is
determined by requiring that the probability of finding these
three valons in a proton be one, i.\ e.
\begin{eqnarray}
\int^1_0 dy_1 \int^{1 - y_1}_0dy_2 \int^{1 -
y_1-y_2}_0dy_3 \, G_{UUD}(y_1, y_2, y_3) = 1 .
\label{2.2}
\end{eqnarray}
Thus we have
\begin{eqnarray}
g = \left[B (\alpha + 1, \beta +1) B(\alpha + 1, \alpha +\beta
+2)\right]^{-1},
\label{2.3}
\end{eqnarray}
where $B(m,n)$ is the beta function.  The single valon
distributions are obtained by appropriate integrations
\begin{eqnarray}
G_U(y) = \int dy_2 \int dy_3 G_{UUD} (y, y_2, y_3) = g B
(\alpha + 1,
\beta +1) y^{\alpha} (1 - y)^{\alpha +\beta +1},
\label{2.4}
\end{eqnarray}
\begin{eqnarray}
G_D(y) = \int dy_1 \int dy_2 G_{UUD} (y_1, y_2, y)= g B
(\alpha + 1, \alpha +1) y^{\beta} (1 - y)^{2\alpha +1}.
\label{2.5}
\end{eqnarray}
The two-valon distributions are trivial because of the
$\delta$-function in Eq.\ (\ref{2.1}).

In a deep inelastic scattering the structure of the proton is
probed to reveal the parton distributions, which in the valon
model are convolutions of $G_v(y)$ with the evolution
functions that describe the valon structure.  The latter have
two varieties:  $K(z,Q^2)$ for the favored partons
and  $L(z,Q^2)$ for the unfavored partons.  That
is, $u$ is favored in $U$ and $d$ is favored in $D$, but they
are unfavored in $D$ and $U$, respectively.  At high $Q^2$
we may ignore the influence of the spectator valons when
one valon is probed (the usual impulse approximation), so
we can write for the $u$ and $d$ quark distribution
functions as
\begin{eqnarray}
x \, u (x,Q^2) = \int^1_x dy \left[ 2G_U (y)
K(x/y,Q^2)
 + G_D(y) L (x/y,Q^2)\right],
\label{2.6}
\end{eqnarray}
\begin{eqnarray}
x \, d (x,Q^2) = \int^1_x dy \left[G_D (y)
K(x/y,Q^2)
 + 2 G_U(y) L (x/y,Q^2)\right].
\label{2.7}
\end{eqnarray}
It should be emphasized that $u(x)$, $d(x)$ and $G(y)$ are
noninvariant distributions defined in the phase space $dx$
and $dy$, while $K(z)$ and $L(z)$ are invariant
distributions defined in the phase space $dz/z$.

The favored distribution $K(z,Q^2 )$ has two
parts, valence and sea, while the unfavored distribution
$L(z,Q^2 )$ has only sea.  The valence part is also
referred to as the non-singlet component,
$K_{NS}(z,Q^2 )$, so we may write
\begin{eqnarray}
K(z,Q^2 ) = K_{NS}(z,Q^2 )
+ L(z,Q^2 ).
\label{2.8}
\end{eqnarray}
At high $Q^2$ both $K_{NS}$ and $L$ can separately be
determined in perturbative QCD.  However, since low-$Q^2$
parton distributions are now available by extrapolation, and
especially since we shall apply the valon model to low-$p_T$
processes (and therefore at low $Q^2$), we use
phenomenological forms for $K_{NS}$ and $L$ at low $Q^2$
with parameters to be determined by the low-$Q^2$ parton
distributions.  We adopt the forms
\begin{eqnarray}
K_{NS} (z) = z^a (1-z)^b/B(a, b + 1),
\label{2.9}
\end{eqnarray}
\begin{eqnarray}
L (z) = \ell _o (1-z)^5 ,
\label{2.10}
\end{eqnarray}
where Eq.\ (\ref{2.9}) satisfies the requirement that there is
only one valence quark in a valon, i.\ e.\,
\begin{eqnarray}
\int^1_0 {dz \over z} K_{NS} (z) = 1 .
\label{2.11}
\end{eqnarray}
Eq.\ (\ref{2.10}) has the usual sea-quark distribution, also
used previously \cite{hz}.

The parton distributions that we use to fit are the ones
determined by the CTEQ collaboration \cite{cteq}.  In
particular, they have the distributions at low $Q^2$,
labeled CTEQ4LQ, posted on the web \cite{cteq4}.  We
choose the one at $Q^2 = 1 \, GeV^2$ evolved from $Q^2_0
= 0.49  \, GeV^2$.  We fit the $u(x,Q^2 )$ and
$d(x,Q^2 )$ distribution functions using Eqs.
(\ref{2.4})-(\ref{2.10}) by varying $\alpha, \beta, a, b$ and
$\ell _0$.  The results are shown in Fig. 1, where the CTEQ
functions are in solid lines, and our fitted curves are in
dotted lines.  We have attempted to fit the $u$-quark
distribution as perfectly as possible; the fit of the $d$-quark
distribution turns out to be good only at high $x$.  Missing
the normalization of $d(x,Q^2 )$ at low $x$ is a
blemish, but is acceptable since we shall use the valence
quark distributions mainly in the large-$x$ region.  Besides,
the reliability of an extrapolation of high-$Q^2$ deep
inelastic scattering data to low $Q^2$ can always be called
into question.

The parameters of the fit are
\begin{eqnarray}
\alpha = 0.70, \quad \beta = 0.25,
\label{2.12}
\end{eqnarray}
\begin{eqnarray}
a = 0.79, \quad b = -0.26, \quad \ell_0 = 0.083 .
\label{2.13}
\end{eqnarray}
The values of $\alpha$ and $\beta$ are not very different
from the ones determined in Ref. \cite{rch4} based on
muon-scattering data at $Q^2 = 22.5 \, GeV^2$ analyzed by
Duke and Roberts \cite{dr}; there we had $\alpha = 0.65$
and $\beta = 0.35$.  As can be seen from Eqs.\ (\ref{2.4})
and (\ref{2.5}), our present result means that $G_U (y)
\propto (1-y)^{1.95}$ and $G_D (y) \propto (1-y)^{2.4}$ at
large $y$, as opposed to $(1-y)^{2.0}$ and $(1-y)^{2.3}$,
respectively, for the previous result.  The values of $a$
and $b$ in Eq.\ (\ref{2.13})
 suggest that $K_{NS}(z,Q^2 )$ is highly peaked
near $z = 1$, according to Eq.\ (\ref{2.9}).  That is as it
should be, since $Q^2$ is only about twice the value of
$Q^2_0$ and thus not much evolution.  If there were no
evolution, i.\ e., $Q^2 = Q^2_0$, then $K_{NS}(z,Q^2_0
)$ would be $\delta(z - 1)$ and $L(z) = 0$, a situation
indicating no probing of the valons.  What we have at $Q^2
= 1 \, GeV^2$ gives only a modest degree of resolution of the
valon structure, and $K_{NS}(z,Q^2_0)$ is
changed from $\delta(z - 1)$ to $(1 - z)^{-0.26}$ in the large
$y$ region, while $L(z,Q^2_0)$ becomes
non-vanishing.  These are the valence and sea quark
distributions that we shall use for nucleon production  at
low $p_T$ in the following.

\section{Momentum Degradation in the  Valon Model}

The valon model for inclusive reactions is basically a
$s$-channel description of particle production in
contradistinction from the Regge-Gribov approach \cite{vng}
which is essentially based on cutting $t$-channel exchanges
in elastic amplitudes.  The two approaches are roughly
complementary in that the former can best describe the
fragmentation region, while the latter is more suitable for the
central region.

It has been known for a long time that the pion inclusive
cross section in the proton fragmentation region has a $x_F$
dependence that is very similar to the quark distribution in
the proton \cite{wo}.  That similarity has been reconfirmed
more recently in $pA$ collisions by the E910 experiment
\cite{bc} even at the relatively low energy of
$E_{\mbox{lab}} = 17.5 GeV$.  It suggests that the
proton structure is highly relevant to the spectra of
particles produced in the fragmentation region.  The
connection between the quark distribution and the pion
inclusive distribution was first put on a calculable basis
by the recombination model
\cite{dh}, which was subsequently improved in the
framework of the valon model
\cite{rch5}.  The idea of recombination as a basis for
hadronization in hadronic collisions is eminently reasonable,
since for a pion to be detected at $x_F = 0.8$, say, it is far
less costly to have a quark at $x_1$ coalescing with an
antiquark at $x_2$, each $<0.8$, to form a pion at $x_F =
x_1 + x_2$,  as compared to the fragmentation process for
which a quark or diquark must first have $x > 0.8$.  In quark jets,
fragmentation is reasonable because hadronization is
initiated by a leading quark, but it is less persuasive when
adapted to hadronic fragmentation unless the projectile
momentum resides entirely in a quark or diquark
\cite{agp,cstt}, contrary to Ochs' observation of the
relevance of the canonical quark distributions.

The $s$-channel treatment of relating quark distributions to
inclusive hadron distributions in $pp$ collisions essentially
regards the effect of the opposite-going initial proton as
unimportant, an approximation that can only be justified in
the fragmentation regions due to short-range correlation in
rapidity.  Thus the valon model, as it has been developed
up to now, is not expected to be applicable to the central
regions, where the interaction between the two incident
particles is of paramount importance.  For $pA$ collisions
even the factorization of the fragmentation properties is not
entirely valid, since the hadron distributions are known to
depend on centrality or target size.  This is the problem that
we shall treat in this paper in the framework of the valon
model.  More specifically, we shall consider the problem of
momentum degradation of the valons due to the nuclear
medium.  Recombination occurs outside the nucleus and is
therefore unaffected.

The NA49 data on $p$-$Pb$ collisions are presented in
terms of two mean values, $\bar{\nu}$, of the number of
participating nucleons in the target: $\bar{\nu} = 6.3$ for
central collisions and $\bar{\nu} = 3.1$ for non-central
collisions \cite{bc}.  We assume a Poissonian fluctuation
from that mean with the distribution
\begin{eqnarray}
P_{\bar{\nu}}(\nu) = {\bar{\nu}^{\nu} \over \nu!}(
e^{\bar{\nu}} - 1)^{-1} ,
\label{3.1}
\end{eqnarray}
which is normalized by
\begin{eqnarray}
\sum^{\infty}_{\nu = 1}P_{\bar{\nu}}(\nu) = 1 ,
\label{3.2}
\end{eqnarray}
where we have excluded the $\nu = 0$ term, since it is
necessary for $\nu \geq 1$ in order to have a collision.
Thus $\nu$ is the number of nucleons in the nucleus that
suffer inelastic collisions in any given event, $\nu$ being an
integer.  That is the counting on the target side, while on the
projectile side we count in terms of the valons.  Let the $i$th
valon encounter $\nu_i$  collisions.  In general, the total
valonic collisions is bounded by
\begin{eqnarray}
 \nu \leq  \sum^3_{i =  1} \nu_i \leq 3\nu.
\label{3.3}
\end{eqnarray}
The upper bound occurs only when all three valons
participate in each of the struck nucleon, while the lower
limit is for only one valon per struck nucleon.  In the
Appendix we shall show that the data favor the lower
bounds, so for simplicity we proceed in the following with
the assumption
\begin{eqnarray}
\nu = \nu_1 + \nu_2 + \nu_3.
\label{3.4}
\end{eqnarray}
It is useful to interpret this in the $t$-channel picture.  The
incident proton is represented by three constituents, each of
which exchanges $\nu_i$ ladders with $\nu_i$ target
nucleons so that the overall diagram is highly non-planar.
Cutting the ladders gives rise to the particles produced in the
$s$ channel, mostly in the central region.

Focusing on the evolution of the valons, we first assume that
the three valons interact with the target independently, since
they are loosely bound to form the proton, just as the
nucleons are in a deuteron.  If we denote the degradation
effect of the nucleus on the $i$th valon by $D(z_i,
\nu_i)$, then we can write the evolution equation on
the valon distribution as
\begin{eqnarray}
y'_1y'_2y'_3G'(y'_1, y'_2, y'_3;\nu_1,
\nu_2, \nu_3) = \int  dy_1 dy_2 dy_3 \,
G(y_1, y_2, y_3)
D({y'_1 \over y_1}, \nu_1)D({y'_2 \over
y_2}, \nu_2)D({y'_3 \over y_3}, \nu_3)
\label{3.5}
\end{eqnarray}
where $G'$ is the valon distribution function after
$(\nu_1, \nu_2, \nu_3)$ interactions with the
nucleus.  As in Eqs.\ (\ref{2.6}) and (\ref{2.7}), $G$ and
$G'$ are noninvariant distributions defined in the phase
space $dy_1\, dy_2\, dy_3$, etc., while
$D(z_i,\nu_i)$ is an invariant distribution defined
in $dz_i/z_i$.  We postpone our discussion of what
$D(z_i,\nu_i)$ is until Sec.\ V.

For an event with $\nu$ collisions the evolved valon
distribution function $G'_{\nu}
(y'_1, y'_2, y'_3)$ with $\nu$ partitioned as in
Eq.\ (\ref{3.4}) is given by the multinomial formula
\begin{eqnarray}
G'_{\nu}(y'_1, y'_2, y'_3) = {1 \over
3^{\nu}} \sum_{[\nu_i]}{\nu ! \over \nu_1 ! \nu_2 ! \nu_3 !}
G'(y'_1, y'_2, y'_3;\nu_1,
\nu_2, \nu_3) ,
\label{3.6}
\end{eqnarray}
where $[\nu_i]$ implies that the summation is over
$\nu_1, \nu_2$, and $\nu_3$ subject to the constraint of
Eq.\ (\ref{3.4}).  Note that if the nucleus had no effect on the
valons, i.e.,
\begin{eqnarray}
D(z_i,\nu_i) = \delta (z_i - 1)
\label{3.7}
\end{eqnarray}
independent of $\nu_i$, then
$G'(y'_1, y'_2, y'_3;\nu_1,
\nu_2, \nu_3)$ becomes
$G(y_1, y_2, y_3)$ in Eq.\ (\ref{3.5}) and so
does $G'_{\nu}(y'_1, y'_2, y'_3)$ in Eq.\
(\ref{3.6}), as it should.  To relate to the experimental
$\bar{\nu}$, we have
\begin{eqnarray}
G'_{\bar{\nu}}(y'_1, y'_2, y'_3) =
\sum_{\nu}
G'_{\nu}(y'_1, y'_2, y'_3)P_{\bar{\nu}}(\nu) ,
\label{3.8}
\end{eqnarray}
where $P_{\bar{\nu}}(\nu)$ is given in Eq.\ (\ref{3.1}).

It is useful to introduce the notion of an effective nucleon
after $\nu$ collisions by giving it a momentum fraction
$y'$ and defining the probability of finding it at  $y'$
by
\begin{eqnarray}
{\cal G}'_{\nu} (y') = \int dy'_1
dy'_2 dy'_3 \,G'_{\nu}
(y'_1,y'_2,y'_3) \, \delta
(y'_1+y'_2+y'_3-y')
\label{3.9}
\end{eqnarray}
Conservation of baryon number requires that
\begin{eqnarray}
\int^1_0 \, dy'{\cal G}'_{\nu} (y' ) = 1 = \int dy'_1
dy'_2  dy'_3 \, G'_{\nu}
(y'_1, y'_2, y'_3).
\label{3.10}
\end{eqnarray}
The possibility that flavor  can change is a secondary issue
that will be discussed later; here, the issue is that the baryon
should be somewhere in the interval $0 \leq y' \leq 1$.  The
second half of Eq.\ (\ref{3.10}) puts a constraint on
$D(z_i,\nu _i)$, since we can obtain from Eqs.\ (\ref{3.5})
and (\ref{2.2})
\begin{eqnarray}
\int {dz \over z} D(z,\nu _i) = 1.
\label{3.11}
\end{eqnarray}
The likelihood that the three evolved valons will
in reality reconstitute a nucleon is extremely low, but the
fictitious nucleon that they form carries a baryon
number that is conserved, and a momentum that is not
conserved.  Indeed, we expect the average $y'$ to decrease
with $\nu$, i.\ e.,
\begin{eqnarray}
\bar{y}'_{\nu} = \int^1_0 dy' y' {\cal G}'_{\nu} (y') < 1.
\label{3.12}
\end{eqnarray}
That is commonly referred to as stopping.  In Sec.\ V we
shall infer from the data what the stopping power is.

Even without stopping, such as in $pp$ collisions, it does not
mean that the real proton produced cannot have $x_F < 1$.
It is known that in $pp$ collisions the proton inclusive cross
section $d \sigma / dx_F$ is nearly flat in $x_F$.  Stopping
goes on top of that distribution, making it roughly
exponential decrease in $x_F$.  How to proceed from the
valon distribution $G'$ to the detected proton distribution
$H_p$ is the subject of the next section.

The convolution equation (\ref{3.5}) can be simplified when
expressed in terms of the moments on account of the
convolution theorem.  Thus let us define
\begin{eqnarray}
\tilde{D}(n_i,\nu_i) = \int^1_0 {dz_i \over z_i} z_i^{n_i-1}
D(z_i,\nu_i),
\label{3.13}
\end{eqnarray}
\begin{eqnarray}
\tilde{G}(n_1, n_2, n_3) = \int^1_0 dy_1 \int^{1-y_1}_0
dy_2 \int^{1-y_1-y_2}_0 dy_3
 \left[ \prod ^3_{i=1}  y_i^{n_i-1} \right] G(y_1, y_2, y_3)
\label{3.14}
\end{eqnarray}
and similarly for $\tilde{G}'(n_1, n_2, n_3; \nu_1, \nu_2,
\nu_3)$ and $\tilde{G}'_{\nu}(n_1, n_2, n_3)$.  It then
follows from Eqs.\ (\ref{3.5}) and (\ref{3.6}) that
\begin{eqnarray}
\tilde{G}'_{\nu}(n_1, n_2, n_3) = {1 \over 3^{\nu}}
\sum_{[\nu_i]} {\nu ! \over \nu_1 !, \nu_2 !, \nu_3
!} \, \tilde{G}(n_1, n_2, n_3) \prod
^3_{i=1} \tilde{D}(n_i,\nu_i).
\label{3.15}
\end{eqnarray}

Now, $D (n_i,\nu_i)$ itself can be described by a
convolution equation \cite{hz}.  If instead of the discrete
$\nu_i$ we use a continuous variable $L$ that denotes the
length of the nuclear medium a valon traverses, we can
express the change on $D(z, L)$ for an incremental distance
$dL$ in the form \cite{hpp}
\begin{eqnarray}
{d \over dL} D(z, L) = \int^1_z {dz' \over z'} D(z', L) Q
(z/z')
\label{3.16}
\end{eqnarray}
with some reasonable kernel $Q (z/z')$.  In terms of the
moments with
\begin{eqnarray}
\tilde{Q}(n) = \int^1_0 d\zeta \, \zeta^{n-2} Q(\zeta) ,
\label{3.17}
\end{eqnarray}
Eq.\ (\ref{3.16}) becomes
\begin{eqnarray}
{d \over dL} \tilde{D}(n, L) = \tilde{D}(n, L)\tilde{Q}(n),
\label{3.18}
\end{eqnarray}
whose solution is
\begin{eqnarray}
\tilde{D}(n, L) = \exp \left[\tilde{Q}(n)L\right].
\label{3.19}
\end{eqnarray}
The constraint (\ref{3.11}) implies
\begin{eqnarray}
\tilde{D}(1, L) = 1 , \quad \mbox{and} \quad \tilde{Q}(1)
= 0 .
\label{3.20}
\end{eqnarray}
Since $\nu_i$ is proportional to $L$, let us now revert
$\tilde{D}(n, L)$ to $\tilde{D}(n_i, \nu_i)$ and write Eq.\
(\ref{3.19}) as
\begin{eqnarray}
\tilde{D}(n_i, \nu_i) = d(n_i)^{\nu_i} ,
\label{3.21}
\end{eqnarray}
where $d(n_i)$ is trivially related to $e^{\tilde{Q}(n_i)}$
with a power exponent whose detail need not be specified
here.  From Eq.\ (\ref{3.20}) follows
\begin{eqnarray}
d(1) = 1.
\label{3.22}
\end{eqnarray}

We now can use Eq.\ (\ref{3.21}) in (\ref{3.15}) and obtain
\begin{eqnarray}
\tilde{G}'_{\nu}(n_1, n_2, n_3) = \tilde{G}(n_1, n_2, n_3)
\left[{1 \over 3} \sum^3_{i = 1} d(n_i)\right]^{\nu} .
\label{3.23}
\end{eqnarray}
What we have derived here is that the dependence on $\nu$
is in the exponent, implying that the effects of the
successive collisions with the nucleons in the target nucleus
are multiplicative, as is reasonable.  We can now go back to
Eq.\ (\ref{3.12}) and calculate the average $\bar{y}'_{\nu}$
after $\nu$ collisions.  Using Eq.\ (\ref{3.9}) we get
\begin{eqnarray}
\bar{y}'_{\nu} &=& \tilde{G}'_{\nu}(2, 1, 1) +
\tilde{G}'_{\nu}(1, 2, 1) + \tilde{G}'_{\nu}(1, 1,
2)\nonumber\\
&=&\left[ \tilde{G}(2, 1, 1) +
\tilde{G}(1, 2, 1) + \tilde{G}(1, 1,2)\right] \left\{[2 + d(2)]
/3\right\}^{\nu}
\label{3.24}
\end{eqnarray}
with the help of Eqs.\ (\ref{3.22}) and (\ref{3.23}).  The first
factor in the square brackets is just 1, since it is
\begin{eqnarray}
\int dy_1 dy_2 dy_3 (y_1 + y_2 + y_3) G (y_1, y_2, y_3)
 = \left< \sum_i y_i\right> = 1 ,
\label{3.25}
\end{eqnarray}
which is guaranteed by the $\delta$-function in Eq.\
(\ref{2.1}).  Hence, we have
\begin{eqnarray}
\bar{y}'_{\nu} = \xi^{\nu}, \quad \quad\xi = [2 +
d(2)]/3 < 1.
\label{3.26}
\end{eqnarray}
Averaging over $P_{\bar{\nu}}(\nu)$, defined in Eq.\
(\ref{3.1}), yields
\begin{eqnarray}
\left<  y'\right>_{\bar{\nu}} = \sum^{\infty}_{\nu = 1}
\bar{y}'_{\nu}P_{\bar{\nu}}(\nu) = (e^{\xi
\bar{\nu}}-1)/(e^{\bar{\nu}}-1) .
\label{3.27}
\end{eqnarray}
This is the average momentum fraction of the effective
nucleon after $\bar{\nu}$ collisions but before
fragmentation into final-state particles in the fragmentation
region.

\section{Fragmentation and Recombination}

We now consider the problem of how a projectile proton
fragments and how the quarks recombine to form the
detected nucleon, thereby specifying how the inclusive
distribution can be calculated.  To give an overview of
the procedure, let us summarize the two steps above by
the two invariant distributions:  $F(x_1, x_2, x_3)$, the
probability of finding a $u$ quark at $x_1$, another
$u$ quark at $x_2$, and a $d$ quark at $x_3$, and
$R_p(x_1, x_2, x_3, x)$, the recombination function,
which specifies the probability that those three quarks
coalesce to form a proton at $x$.  How these
distributions are related to the valon distributions will
be discussed later.  But first we state that the invariant
distribution function for the detection of a proton at
$x$ is
\begin{eqnarray}
{x  \over  \sigma_{in}}{d\sigma^p  \over  dx}  \equiv H_p
(x) = {1 \over N} \int {dx_1 \over x_1}{dx_2 \over
x_2}{dx_3 \over x_3} F(x_1, x_2, x_3)R_p(x_1, x_2, x_3,
x)
\label{4.1}
\end{eqnarray}
where the normalization factor $N$ will be given below.
Eq.\ (\ref{4.1}) is the essence of the recombination model
\cite{rch5}-\cite{hz}.  For meson production, only the
distributions for $q$ and $\bar{q}$ need be considered and
Eq.\ (\ref{4.1}) can be simplified accordingly.  As in all
distributions considered in this paper, color and spin
components are averaged over in the initial state and
summed in the final state so that we are not concerned
explicitly with such degrees of freedom.  Flavor, however, is
different, since we identify the final-state particles by their
flavors; that problem will be treated presently.

Before proceeding we emphasize what has been mentioned in
the preceding section already, namely:  Eq.\ (\ref{4.1}) is
expected to be valid in the proton fragmentation region
only, if $F(x_1, x_2, x_3)$ is to be determined from the
projectile valon distributions with no quarks originating
from the target nucleons.  In this $s$-channel approach, the
factorization of the projectile and target fragmentations,
apart from the momentum-degradation effect studied in the
previous section, can be justified only if the two
fragmentation regions are well separated.  At AGS that is
not the case.  Even at SPS the central region in rapidity can
encompass sizable portions of the positive and negative
$x_F$ variables.  The application of the valon model to the
analysis of the data therefore needs some help from the
experiments.

Fortunately, the NA49 collaboration has treated their data in
such a way as to eliminate the contribution of the target
fragmentation  from the projectile fragmentation region.
From their  data on the net proton produced, $p + A
\rightarrow (p - \bar{p}) + X$, for which we use the
abbreviated notation $(p - \bar{p})_p$, they subtract the
distribution for $(p - \bar{p})_{\pi}$, which is ${1 \over
2}[(p - \bar{p})_{\pi^+} + (p - \bar{p})_{\pi^-}]$.  By charge
conjugation symmetry, $(p - \bar{p})_{\pi}$ should have
no projectile fragmentation, only target fragmentation.
Thus the difference $(p - \bar{p})_p - (p - \bar{p})_{\pi}$
should have no target fragmentation \cite{gv}.  With those
data as our goal for analysis, Eq.\ (\ref{4.1}) is then
particularly suitable.

The normalization factor $N$ in Eq.\ (\ref{4.1}) is
determined by requiring that the proton distribution in the
absence of any target, $H^0_p(x)$, satisfies the sum rules
\begin{eqnarray}
\int^1_0 {dx \over x} H^0_p(x) = \int^1_0 dx  H^0_p(x) =
1 ,
\label{4.2}
\end{eqnarray}
which follow from the condition that the number of proton
and its momentum fraction be 1.  Without any collision the
quarks are identified with the unresolved valons, so $F(x_1,
x_2, x_3)$ becomes
\begin{eqnarray}
F^0(x_1, x_2, x_3) =  x_1 x_2 x_3 G_{UUD}(x_1, x_2,
x_3) .
\label{4.3}
\end{eqnarray}
The recombination function is the time-reversed form of
the valon distribution, so
\begin{eqnarray}
R_p (x_1, x_2, x_3, x) =  {x_1 x_2 x_3 \over
x^3}G_{UUD}\left({x_1 \over x}, {x_2 \over x}, {x_3
\over x}\right).
\label{4.4}
\end{eqnarray}
Putting these in Eq.\ (\ref{4.1}), we obtain in view of
(\ref{2.1})
\begin{eqnarray}
H^0_p(x) = {g^2 \over N} \int^1_0 dx_1 \int^{1-x_1}_0
dx_2 (x_1 x_2)^{2 \alpha + 1} (1-x_1- x_2)^{2 \beta + 1}
x^{-(2 \alpha + \beta + 2)} \delta(x-1).
\label{4.5}
\end{eqnarray}
Because of the presence of $\delta(x-1)$, the two integrals
in Eq.\ (\ref{4.2}) are identical, and we get
\begin{eqnarray}
N = g^2B (2 \alpha +2, 2 \alpha + 2 \beta +4) B(2 \alpha +
2, 2 \beta + 2).
\label{4.6}
\end{eqnarray}
The factor $g^2$, although known from Eq.\ (\ref{2.3}),
 will cancel the similar factor that will emerge from the
integral in the numerator of Eq.\ (\ref{4.1}), just as they
appear explicitly in Eq.\ (\ref{4.5}).

The identification of the recombination function with the
invariant form of the valon distribution in Eq.\ (\ref{4.4}) is
the principle characteristic of the valon model.  On the one
hand, it recognizes the role of the wave function of the
proton both in a projectile and in a produced proton.  On
the other hand, the momentum fractions $x_i$ of the
outgoing valons can add up to a proton at $x$, so there is no
need for any constituent in the process to have a
momentum fraction greater than $x$, as would be
necessary in a quark fragmentation model.  One may then
question how in a collision process can the quarks at $x_1,
x_2$, and  $x_3$ in Eq.\ (\ref{4.1}) become the valons of
the outgoing proton.  The answer is that hadronization
occurs outside the target, and that the quarks moving
downstream dress up themselves and
become the valons of the produced proton without any
change in the net momentum of each quark/valon, which
is all that matters in the specification of $R_p (x_1,
x_2, x_3, x)$.

We now consider the quark distribution $F(x_1, x_2,
x_3)$ in Eq.\ (\ref{4.1}) in $pA$ collision before
recombination.  In the preceding section we have formulated
the procedure to calculate the effect of the nuclear medium
on the momenta of the valons as they traverse the target.
Momentum degradation is, however, only one of the effects
of valon-nucleon interaction.  If such an interaction is
represented by a Regge exchange, we should also consider
the possibility of flavor changes of the valons due to
non-vacuum exchanges at non-asymptotic energies.  In the
spirit of the $s$-channel approach that we have taken,
in which the probabilities of occurrences at various stages
are assembled multiplicatively, we assume that the flavor
changes at each of the $\nu_i$ collisions are incoherent so
that the net probability of a flavor change after $\nu _i$
interactions is also multiplicative.  Let $q$ be the probability
of a flavor change from $U$ to $D$, or from $D$ to $U$, at
one of the $\nu_i$ interactions.  Furthermore, let
$q_{\nu_i}$ be the probability of flavor change after
$\nu_i$ interactions.  Then $q_{\nu_i}$ satisfies the
recursion relation
\begin{eqnarray}
q_{\nu_i + 1} = q_{\nu_i} (1-q) + (1 - q_{\nu_i}) q,
\label{4.7}
\end{eqnarray}
where the first term on the RHS denotes no change in the
last step from $\nu_i$ to $\nu_{i+1}$, while the second
term denotes a change in the last step.  The solution of
Eq.\ (\ref{4.7}) is
\begin{eqnarray}
q_{\nu_i} = {1 \over 2} [1 - (1-2q)^{\nu_i}].
\label{4.8}
\end{eqnarray}
We may now write what a $U$ and a $D$ valon become
after
$\nu_i$ interactions in obvious notation:
\begin{eqnarray}
U \stackrel{\nu_i}{\rightarrow} p_{\nu_i}U + q_{\nu_i}D
\label{4.9}
\end{eqnarray}
\begin{eqnarray}
D \stackrel{\nu_i}{\rightarrow} p_{\nu_i}D + q_{\nu_i}U
\label{4.10}
\end{eqnarray}
where $p_{\nu_i} = 1 -q_{\nu_i}$.  This regeneration
process depends on one parameter $q$.  We expect $q$ to
decrease with energy $\sqrt{s}$.  Here we treat it as one free
parameter to fit the NA49 data at one energy
$E_{\mbox{lab}} = 158\, GeV$.

For the quark distribution in a valon we have the favored and
unfavored types discussed in Sec. II, and denoted by $K(z)$
and $L(z)$, respectively.  We drop the $Q^2$ dependence,
since we now consider low-$p_T$ hadronic processes for
which there is no precise $Q^2$.  Nevertheless, we shall use
the parameterization in
Eqs.\ (\ref{2.9}), (\ref{2.10}) and (\ref{2.13}), from the
$Q^2 = 1 \, GeV^2$ CTEQ parton distributions.  We use
the following notation for the invariant distributions of
quarks in valons with superscript $f$ signifying `favored'
and $u$ `unfavored':

$V^f_{ij}$:  favored quark at $x_j$ in valon at $y'_i$,

$W^{ff}_{i,jk}$:  two favored quarks at $x_j$ and $x_k$ in
valon at $y'_i$.

\noindent Other distributions involving unfavored quarks
are similarly defined.  Examples  of $V^f_{ij}$,
$V^u_{ij}$,
$W^{ff}_{i,jk}$, and $W^{fu}_{i,jk}$ are depicted by
diagrams in Fig.\ 2.

In view of  Eq.\ (\ref{4.9}) and Eq.\ (\ref{4.10}) we can write
by definition
\begin{eqnarray}
V^f_{ij} =  p_{\nu_i} K\left({x_j \over y'_i}\right) +
q_{\nu_i} L\left({x_j
\over y'_i}\right)
\label{4.11}
\end{eqnarray}
\begin{eqnarray}
V^u_{ij} =  p_{\nu_i} L \left({x_j \over y'_i}\right) +
q_{\nu_i} K\left({x_j \over y'_i}\right)
\label{4.12}
\end{eqnarray}
\begin{eqnarray}
W^{ff}_{i,jk} =  p_{\nu_i} \left\{K \left({x_j \over
y'_i}\right) L\left({x_k
\over y'_i - x_j}\right)\right\}_{jk} +
q_{\nu_i} \left\{L \left({x_j \over y'_i}\right) L\left({x_k
\over y'_i - x_j}\right)\right\}_{jk}
\label{4.13}
\end{eqnarray}
\begin{eqnarray}
W^{uu}_{i,jk} =  p_{\nu_i}
\left\{L \left({x_j \over y'_i}\right) L\left({x_k
\over y'_i - x_j}\right)\right\}_{jk} +
q_{\nu_i}\left\{K \left({x_j \over y'_i}\right) L\left({x_k
\over y'_i - x_j}\right)\right\}_{jk}
\label{4.14}
\end{eqnarray}
\begin{eqnarray}
W^{fu}_{i,jk} =  p_{\nu_i} \left\{K \left({x_j \over
y'_i}\right) L\left({x_k
\over y'_i - x_j}\right)\right\}_{jk} +
q_{\nu_i} \left\{L \left({x_j \over y'_i}\right) K\left({x_k
\over y'_i - x_j}\right)\right\}_{jk}
\label{4.15}
\end{eqnarray}
where
\begin{eqnarray}
\left\{f_1 \left({x_j \over y'_i}\right) f_2 \left({x_k
\over y'_i - x_j}\right) \right\}_{jk} = {1
\over 2}\left[f_1 \left({x_j \over y'_i}\right) f_2 \left({x_k
\over y'_i - x_j}\right)+ f_2 \left({x_k \over
y'_i}\right) f_1 \left({x_j \over
y'_i-x_k}\right)\right] .
\label{4.16}
\end{eqnarray}
In terms of these $V$ and $W$ distributions we can now
write out, by inspection, all possible contributions to the
$uud$ quarks, shown in Fig.\ 3, for the production of a
proton
\begin{eqnarray}
M_p(y'_1,y'_2,y'_3;x_1, x_2, x_3) &=&
V^f_{11}V^f_{22}V^f_{33} + 2
\{V^f_{11}V^{u}_{23}V^u_{32}\}_{12}\nonumber\\
&&+2\{V^f_{11}W^{fu}_{2,23}\}_{12}
+ 2V^u_{13}W^{ff}_{2,12}
+ 2\{V^f_{11}W^{fu}_{3,32}\}_{12}\nonumber\\
&&+ 2V^u_{13}W^{uu}_{3,12}
+2\{V^u_{31}W^{fu}_{1,23}\}_{12}
+ 2V^f_{33}W^{ff}_{1,12} ,
\label{4.17}
\end{eqnarray}
where $\{\,\}_{12}$ denotes symmetrization of $x_1$ and $x_2$.  We have ignored the
contributions corresponding to all three quarks coming from
the same valon.  The quark distribution from the proton source
for proton production is then
\begin{eqnarray}
F_p(x_1, x_2, x_3) = \int dy'_1
dy'_2 dy'_3 \, G'_{\bar{\nu}}
(y'_1, y'_2, y'_3)M_p(y'_1, y'_2, y'_3; x_1, x_2, x_3)
\label{4.18}
\end{eqnarray}
where $G'_{\bar{\nu}} (y'_1, y'_2, y'_3)$ is given in Eq.\
(\ref{3.8}).  For $\bar{p}$ production we need only change all
$K$ functions in  Eqs.\ (\ref{4.11}) - (\ref{4.15}) to $L$
functions, since all anti-quarks are in the sea.  Denoting the
corresponding quantities in Eqs.\ (\ref{4.17}) and (\ref{4.18})
by $M_{\bar{p}}$ and $F_{\bar{p}}$, we have finally for net
proton production
\begin{eqnarray}
F_{p-\bar{p}} = G'_{\bar{\nu}} \otimes (M_p -
M_{\bar{p}}) ,
\label{4.19}
\end{eqnarray}
where the convolution is defined by the integral in Eq.\
(\ref{4.18}).  It should be recognized that
$G'_{\bar{\nu}} (y'_1, y'_2, y'_3)$ involves a summation of
$G' (y'_1, y'_2, y'_3)$ over
$\nu$, which in turn involves a summation
$G' (y'_1, y'_2, y'_3; \nu_1, \nu_2, \nu_3)$
over $\nu_1$, $\nu_2$ and $\nu_3$ that appear in the
$V$ and
$W$ distributions.

We are now ready to substitute $F(x_1, x_2, x_3)$ and $R(x_1,
x_2, x_3, x)$, defined in Eq.\ (\ref{4.4}), into Eq.\
(\ref{4.1}) to calculate $H(x)$.  Nine convolution integrals
are involved:
$y_i, y'_i$, and $x_i$.  Obviously, we should go to the
moments and reduce them to products.  First, using
Eqs.\ (\ref{2.1}) in (\ref{4.4}), we have
\begin{eqnarray}
H(x) = {g \over N} x ^{-(2\alpha + \beta +2)} \int
dx_1dx_2dx_3 F(x_1, x_2, x_3)(x_1  x_2)^{\alpha + 1}
x_3^{\beta +1} \delta (x_1+ x_2+ x_3-x).
\label{4.20}
\end{eqnarray}
For convenience, let us leave out the known factors
and define
\begin{eqnarray}
H'(x) = {N \over g} x ^{2\alpha + \beta +4} H(x).
\label{4.21}
\end{eqnarray}
Then define the moments
\begin{eqnarray}
\tilde{H}'(n) = \int^1_0 dx  x^{n-2} H'(x),
\label{4.22}
\end{eqnarray}
\begin{eqnarray}
\tilde{F}(n_1, n_2, n_3) =
\int dx_1 dx_2 dx_3 \left(\prod ^3_{i = 1} x_i^{n_i-2}\right)
F(x_1, x_2, x_3) .
\label{4.23}
\end{eqnarray}
Thus it follows
\begin{eqnarray}
\tilde{H}'(n) =  \sum_{[n_i]} {n ! \over n_1 ! n_2
! n_3 !} \tilde{F}  (n_1 + \alpha + 3, n_2 + \alpha + 3,
n_3 + \beta + 3) .
\label{4.24}
\end{eqnarray}
From Eq.\ (\ref{4.18}) we expect from the convolution
theorem to have
\begin{eqnarray}
\tilde{F}(n_1, n_2, n_3) = \tilde{G}'_{\bar{\nu}}(n_1,
n_2, n_3) \tilde{M}(n_1, n_2, n_3)
\label{4.25}
\end{eqnarray}
where $\tilde{G}'_{\bar{\nu}}$ is given by Eq.\
(\ref{3.15}), apart from the Poissonian sum of
(\ref{3.8}).  However, because of the ordering of $x_j$
relative to $y'_i$ in Eq.\ (\ref{4.17}), the simple form of
(\ref{4.25}) is valid only for the first term of $M(y'_1,
y'_2, y'_3; x_1, x_2, x_3)$.  For that first term, which we
denote by $\tilde{F}^{(1)}$, we have
\begin{eqnarray}
\tilde{F}^{(1)}(n_1, n_2, n_3) = \tilde{G}'(n_1, n_2,
n_3) \prod ^3_{i = 1} [p_{\nu_i} \tilde{K}(n_i) +
q_{\nu_i}
\tilde{L}(n_i)] ,
\label{4.26}
\end{eqnarray}
where the subscript $\bar{\nu}$ has been omitted.
The summation over $\nu_i$ in Eq.\ (\ref{3.15}) for $\tilde{G}'(n_1, n_2,
n_3)$ should extend over $p_{\nu_i}$ and
$q_{\nu_i}$ in Eq.\ (\ref{4.26}).  The moments of
$K(z_i)$ and
$L(z_i)$ are defined as usual for the invariant
distributions as
\begin{eqnarray}
\tilde{K}(n_i) = \int^1_0 dz_i  z_i^{n_i-2} K(z_i) .
\label{4.27}
\end{eqnarray}
For all other terms in Eq.\ (\ref{4.17}) there are
$V^u_{ij}$ and $W$ functions, and the simple form of
Eq.\ (\ref{4.25}) must be modified, especially when only
two valons contribute.  For notational simplicity let us
define
\begin{eqnarray}
\tilde{V}^f_{ij} = p_{\nu_i} \tilde{K}(n_j) + q_{\nu_i}
\tilde{L}(n_j) ,
\label{4.28}
\end{eqnarray}
\begin{eqnarray}
\tilde{V}^u_{ij} = p_{\nu_i} \tilde{L}(n_j) + q_{\nu_i}
\tilde{K}(n_j) ,
\label{4.29}
\end{eqnarray}
\begin{eqnarray}
\tilde{W}^{ff}_{i,jk} = p_{\nu_i} \{\tilde{K}(n_j, n_k)
\tilde{L}(n_k)\}_{jk} +   q_{\nu_i}\{\tilde{L}(n_j, n_k)
\tilde{L}(n_k)\}_{jk},
\label{4.30}
\end{eqnarray}
\begin{eqnarray}
\tilde{W}^{uu}_{i,jk} = p_{\nu_i} \{\tilde{L}(n_j, n_k)
\tilde{L}(n_k)\}_{jk} +   q_{\nu_j}\{\tilde{K}(n_j, n_k)
\tilde{L}(n_k)\}_{jk},
\label{4.31}
\end{eqnarray}
\begin{eqnarray}
\tilde{W}^{fu}_{i,jk} = p_{\nu_i} \{\tilde{K}(n_j, n_k)
\tilde{L}(n_k)\}_{jk} +   q_{\nu_j}\{\tilde{L}(n_j, n_k)
\tilde{K}(n_k)\}_{jk},
\label{4.32}
\end{eqnarray}
where $\tilde{K}(n_j, n_k)$ is defined by
\begin{eqnarray}
\tilde{K}(n_j, n_k) = \int^1_0 dz  z^{n_j-2}
(1-z)^{n_k-1} K(z)
\label{4.33}
\end{eqnarray}
and similarly for $\tilde{L}(n_j, n_k)$.  These moments
arise whenever two quarks are from the same valon, as they
do for all $W$ distributions.  Take, for example, the third
term in Eq.\ (\ref{4.17}); we have
\begin{eqnarray}
\tilde{F}^{(3)}(n_1, n_2, n_3)
\hspace{-1in}&& \nonumber\\   &=&  \int^1_0 dy'_1
\int^{1-y'_1}_0 dy'_2\,G'_{UU}(y'_1, y'_2)
\int\left[\prod ^3_{i = 1} dx_i x_i^{n_i-2}\right]
2\left\{ p_{\nu_1}p_{\nu_2} K({x_1 \over y'_1})K({x_2
\over y'_2})L({x_3 \over y'_2-x_2}) +
\cdots\right\}\nonumber\\ &=&   \int^1_0 dy'_1
\int^{1-y'_1}_0 dy'_2 G'_{UU} (y'_1, y'_2)
y'^{n_1-1}_1  y'^{n_2+n_3-2}_2 2p_{\nu_1}p_{\nu_2}
\nonumber\\ &&\cdot \int^1_0 dz_1
z_1^{n_1-2}K(z_1)\int^1_0 dz_2
z_2^{n_2-2}(1-z_2)^{n_3-1}K(z_2)
\int^1_0 dz_3 z_3^{n_3-2}L(z_3) +
\cdots \nonumber\\
&=&2 \{\tilde{G}'_{UU} (n_1, n_2 +
n_3-1) \tilde{V}^f_{11}\tilde{W}^{fu}_{2,23}
\}_{12},
\label{4.34}\end{eqnarray}
where $\{ \}_{12}$ here means symmetrization of $n_1$ and
$n_2$, and
\begin{eqnarray}
G'_{UU} (m_1, m_2)= G'_{UUD} (m_1, m_2,
m_3 = 1).
\label{4.34a}
\end{eqnarray}
Performing the same type of operations on
all terms we obtain
\begin{eqnarray}
\tilde{F}_p (n_1, n_2, n_3) \hspace{-1in}
&&  \nonumber\\
&=&G'_{UUD} (n_1, n_2,
n_3)\tilde{V}^f_{11}\tilde{V}^f_{22}\tilde{V}^f_{33} + 2
\{G'_{UUD} (n_1, n_3,
n_2)
\tilde{V}^f_{11}\tilde{V}^u_{32}\}_{12}\tilde{V}^u_{23}\nonumber\\
&&+ 2 \{G'_{UU} (n_1, n_2 +
n_3-1)\tilde{V}^f_{11}\tilde{W}^{fu}_{2,23}
\}_{12}+ 2 G'_{UU} (n_3, n_1 + n_2-1)
\tilde{V}^u_{13}\tilde{W}^{fu}_{2,12}\nonumber\\
&&+ 2 \{G'_{UD} (n_1, n_2 + n_3-1)
\tilde{V}^f_{11}\tilde{W}^{fu}_{3,32}\}_{12}
+2 G'_{UD} (n_3, n_1 + n_2-1)
\tilde{V}^u_{13}\tilde{W}^{uu}_{3,12}\nonumber\\ &&+ 2
\{G'_{UD} (n_2 + n_3 -1, n_1)
\tilde{W}^{fu}_{1,23}\tilde{V}^u_{31}\}_{12} + 2 \{G'_{UD}
(n_1 + n_2 -1, n_3)
\tilde{W}^{ff}_{1,12}\tilde{V}^f_{33}\}_{12}.
\label{4.35}
\end{eqnarray}
Substituting this in Eq.\ (\ref{4.24}), we have the final
form for the moments $\tilde{H}'_p(n)$.  For $\bar{p}$
production we need only change all $\tilde{K}$ in Eqs.\
(\ref{4.28}) - (\ref{4.32}) to $\tilde{L}$.

For the production of neutron the recombination
function is
\begin{eqnarray}
R_n (x_1, x_2, x_3, x) =  {x_1 x_2 x_3 \over
x^3}G_{DDU}\left({x_1 \over x}, {x_2 \over x}, {x_3 \over
x}\right) ,
\label{4.36}
\end{eqnarray}
where we use $y_1$ and $y_2$ to refer to $D$ and $y_3$
to $U$, so the dependence of $G_{DDU}(y_1, y_2, y_3)$
on $y_i$ is the same as that of $G_{UUD}(y_1, y_2, y_3)$
given in Eq.\ (\ref{2.1}).  The quark distribution $F_n(x_1,
x_2, x_3)$ with the corresponding identification of $x_1$
and
$x_2$ with the $d$ quark, and $x_3$ with the $u$ quark,
also has eight terms, as shown in Fig.\ 4.  We can write,
by inspection, the moments $\tilde{F}_n(x_1, x_2,
x_3)$ similar to Eq.\ (\ref{4.35})
\begin{eqnarray}
\tilde{F}_n\hspace{-.25in}&&(x_1, x_2,
x_3)=\nonumber\\
 && 2 \{
\tilde{G}'_{UUD} (n_3, n_1, n_2) \tilde{V}^f_{13}
\tilde{V}^u_{21} \tilde{V}^u_{32}\}_{12}
+\tilde{G}'_{UUD}
(n_1, n_2, n_3) \tilde{V}^u_{11}
\tilde{V}^u_{22} \tilde{V}^u_{33}
\nonumber\\ &&+2 \tilde{G}'_{UU}
(n_3, n_1+ n_2-1) \tilde{V}^f_{13}
\tilde{W}^{uu}_{2, 12}
+2 \{\tilde{G}'_{UU}
(n_1, n_2+ n_3-1) \tilde{V}^u_{11}
\tilde{W}^{fu}_{2, 32} \}_{12}\nonumber\\
&&+2 \tilde{G}'_{UD}
(n_3, n_1+ n_2-1) \tilde{V}^f_{13}
\tilde{W}^{ff}_{3, 12}
+2 \{\tilde{G}'_{UD}
(n_1, n_2+ n_3-1) \tilde{V}^u_{11}
\tilde{W}^{fu}_{3, 23} \}_{12}\nonumber\\
&&+2 \tilde{G}'_{UD}
(n_1+ n_2-1, n_3) \tilde{W}^{uu}_{1,12}
\tilde{V}^{u}_{3, 3}
+2 \{\tilde{G}'_{UD}
(n_1+ n_3-1, n_2) \tilde{W}^{fu}_{1,31}
\tilde{V}^{f}_{3, 2} \}_{12}  .
\label{4.37}
\end{eqnarray}
Substituting this in Eq.\ (\ref{4.24}) gives
$\tilde{H}'_n(n)$.  For $\bar{n}$ production simply
replace all $\tilde{K}$ in $\tilde{H}'_n(n)$ by $\tilde{L}$.

Finally for net nucleon production we have
\begin{eqnarray}
\tilde{H}'_{p-\bar{p}}(n) = \tilde{H}'_p(n) -
\tilde{H}'_{\bar{p}}(n) ,
\label{4.38}
\end{eqnarray}
\begin{eqnarray}
\tilde{H}'_{n-\bar{n}}(n) = \tilde{H}'_n(n) -
\tilde{H}'_{\bar{n}}(n) .
\label{4.389}
\end{eqnarray}
We shall use these in the next section to determine the $x$
dependences to be compared with the data.

\section{Net Proton and Neutron Distributions}

Before we can compute the distributions $H(x)$, we need to
specify quantitatively the degradation function $D(z_i,
v_i)$ introduced in Eq.\ (\ref{3.5}) and discussed between
Eq.\ (\ref{3.16}) and  (\ref{3.22}).  We proposed an
evolution equation for $D(z, L)$ in Eq.\ (\ref{3.16}) but left
the kernel $Q(z/z')$ unspecified.  Now, to proceed we must
specify $Q(\zeta)$, which is uncalculable because it
represents the non-perturbative effect of the nuclear
medium on a valon as it propagates an incremental
distance.  We shall use an one-parameter description of the
effect, so we shall be approximate by assuming that the
effect is like a one-gluon exchange \cite{hz}, i.\ e.,
\begin{eqnarray}
Q(\zeta) = {\kappa\zeta \over (1 - \zeta)_+},
\label{5.1}
\end{eqnarray}
where the singularity at $\zeta = 1$ is regularized by the
subtraction
\begin{eqnarray}
{1 \over (1 - \zeta)_+} = {1 \over 1 - \zeta} - \delta (\zeta -
1) \int^1_0 {dx \over 1 - x} .
\label{5.2}
\end{eqnarray}
Evidently, Eq.\ (\ref{5.1}) satisfies the condition
\begin{eqnarray}
\int^1_0 {d\zeta \over \zeta} Q(\zeta) = 0 ,
\label{5.3}
\end{eqnarray}
which is required by the constraint $\tilde{Q}(1) = 0$
stated in Eq.\ (\ref{3.20}) that follows from baryon
conservation.  From the definition of the moments given in
Eq.\ (\ref{3.17}) we find, using Eq.\ (\ref{5.1}),
\begin{eqnarray}
\tilde{Q}(n) = - \kappa \sum^{n-1}_{j=1} =  - \kappa
[\psi(n) + \gamma_E],
\label{5.4}
\end{eqnarray}
where $\psi(n)$ is the digamma function and $\gamma_E$
is the Euler's constant, 0.5722.  Substituting $\tilde{Q}(n)$
into Eq.\ (\ref{3.19}), and changing $L$ to $\nu _i$ that
involves a constant factor, thereby effecting a change from
one unknown parameter $\kappa$ to another, $k$, we
obtain the form for $\tilde{D}(n_i, \nu_i)$ in Eq.\
(\ref{3.21}) with
\begin{eqnarray}
d(n_i) = \exp \{- k  [\psi(n_i) + \gamma_E]\}.
\label{5.5}
\end{eqnarray}
This is a one-parameter description of the effect of
momentum degradation.  We shall vary $k$ to fit the data.
Eq.\ (\ref{5.5}) is a rigorous consequence of a simple form
for $Q(\zeta)$ given in Eq.\ (\ref{5.1}), whose reliability is
unknown.  The validity of $d(n_i)$ as expressed in Eq.\
(\ref{5.5}) can only be inferred {\it a posteriori}
 from the fit of the data.  The exponential dependence on
 the degradation strength $k$ follows only from the
linear dependence of $Q(\zeta)$ on $\kappa$, and is
sensible.

We have only two free parameters, $k$ and $q$, to vary to
fit the NA49 data on $p$-$\bar{p}$ and $n$-$\bar{n}$
\cite{bc}.  Recall that $q$ is introduced in Eq.\ (\ref{4.7})
in connection with flavor changes.  We repeat that the data
do not include target fragmentation because
$(p-\bar{p})_{\pi}$ and $(n-\bar{n})_{\pi}$  have
been subtracted out.  Thus the data represent only
proton fragmentation and are ideal for our analysis by
$H_{p-\bar{p}}(x)$ and
$H_{n-\bar{n}}(x)$.

Since the data are in the $x$ variable, we must make the
inverse transformation from our moments to $H(x)$.
Instead of making the inverse Mellin transform, let us
exploit the orthogonality of the Legendre polynomials and
shift the variable to the interval $0 \leq x  \leq 1$.  Thus,
define
\begin{eqnarray}
g_{\ell}(x) = P_{\ell}(2x - 1)
\label{5.6}
\end{eqnarray}
so that
\begin{eqnarray}
\int^1_0 dx g_{\ell}(x)g_m(x) = { 1 \over 2\ell +1}
\delta_{\ell m}.
\label{5.7}
\end{eqnarray}
If we expand the distribution $H'(x)$ in terms of
$g_{\ell}(x)$
\begin{eqnarray}
H'(x) = \sum^{\infty}_{\ell = 0} (2\ell
+1)h_{\ell}g_{\ell}(x),
\label{5.8}
\end{eqnarray}
then the inverse is
\begin{eqnarray}
h_{\ell} = \int^1_0 dx H'(x)g_{\ell}(x).
\label{5.9}
\end{eqnarray}
These $h_{\ell}$ can be expressed in terms of the moments
$H'(n)$ if we express $g_{\ell}(x)$ as a power series in $x$
\begin{eqnarray}
g_{\ell}(x) = \sum^{\ell}_{i = 0}a^i_{\ell}x^i ,
\label{5.10}
\end{eqnarray}
where $a^i_{\ell}$ are known from the properties of
$P_{\ell}(z)$.  Thus from Eq.\ (\ref{5.9}) we have
\begin{eqnarray}
h_{\ell} = \sum^{\ell}_{i = 0}a^i_{\ell} \tilde{H}'(i+2) ,
\label{5.11}
\end{eqnarray}
where $\tilde{H}'(n)$ is defined in Eq.\ (\ref{4.22}).  It is
now clear that our theoretical results in $\tilde{H}'(n)$ can
be transformed to $\tilde{H}'(x)$ through Eqs.\
(\ref{5.8}) and (\ref{5.11}) once we have the coefficients
$a^i_{\ell}$.  Furthermore, if $\tilde{H}'(n)$ becomes
unimportant for $n > N$, then the sum in Eq.\ (\ref{5.8})
can terminate at $N$.

To determine $a^i_{\ell}$, we make use of the recursion
formula
\begin{eqnarray}
(\ell + 1) P_{\ell + 1}(z) = (2\ell + 1) z P_{\ell}(z) -
\ell P_{\ell - 1}(z)
\label{5.12}
\end{eqnarray}
to infer through Eqs.\ (\ref{5.6}) and (\ref{5.10})
\begin{eqnarray}
a^0_{\ell} = -{1 \over \ell}[(2 \ell-1) a^0_{\ell -1} +
(\ell-1) a^0_{\ell -2}],
\label{5.13}
\end{eqnarray}
\begin{eqnarray}
a^i_{\ell} = -{1 \over \ell}[(2 \ell-1) (a^i_{\ell -1} -
2a^{i-1}_{\ell -1} )+ (\ell-1) a^i_{\ell -2}],
\label{5.14}
\end{eqnarray}
where $\ell \geq 2$, and $1 \leq i \leq \ell$.  For $i = 0$
and/or $\ell < 2$, we have
\begin{eqnarray}
a^0_0 = 1, \qquad a^0_1 = -1, \qquad a^1_1 = 2 .
\label{5.15}
\end{eqnarray}
With these we can generate all $a^i_{\ell}$, so $h_{\ell}$
can be directly computed.

Summarizing our procedure, we calculate
$\tilde{H}'_{p-\bar{p}}(n)$ and
$\tilde{H}'_{n-\bar{n}}(n)$, substitute them in Eq.\
(\ref{5.11}) and then (\ref{5.8}), and then use Eq.\
(\ref{4.21}) to determine $H_{p-\bar{p}}(x)$ and
$H_{n-\bar{n}}(x)$.  We vary $k$ and $q$ to fit the data of
NA49 shown in Fig.\ 5.  The inclusive distribution
$dN/dx_F$ corresponds to our $H(x)/x$.  The solid lines
are our results for $p-\bar{p}$ and the dashed line
$n-\bar{n}$ for both $\bar{\nu}= 3.1$ and $6.3$.  The
values of the parameters adjusted are
\begin{eqnarray}
k = 0.62, \qquad \qquad q = 0.37 .
\label{5.16}
\end{eqnarray}
The most striking aspect of our result is that the
normalization of the calculated distributions turns out to
be correct, even though we have no free parameter to
adjust that.  The degradation strength $k$ affects the shape
of the distributions and the flavor-flip probability $q$
affects the difference between $p-\bar{p}$ and
$n-\bar{n}$.  The agreement between theory and
experiment is fairly good, considering that we have
only two free parameters and that the experimental
errors (a typical size of which is shown in the figure)
are large, especially at large $x$.  The shapes of the
distributions for
$p-\bar{p}$ are reasonably well reproduced and so is
the dependence on $\bar{\nu}$.  For $n-\bar{n}$
the calculated curves are somewhat steeper than the
data.  However, there exist some data points for
$n-\bar{n}$ above $x = 0.8$ that are significantly lower,
though with much larger errors.  Taken as a whole the
agreement is satisfactory.  Thus we conclude that the
physical process of proton fragmentation and the nucleon
momentum spectra are well understood in the framework of
the valon model.

We can determine from the value of $k$ what the
momentum degradation length is.  First, we get from Eq.\
(\ref{5.5})
\begin{eqnarray}
d(2) = e^{-k} = 0.54
\label{5.17}
\end{eqnarray}
for $k = 0.62$, since $\psi(2) + \gamma_E = 1$.  By
definition Eq.\ (\ref{3.26}) we obtain $\xi = 0.85$.
Inserting this in Eq.\ (\ref{3.27}) we obtain a dependence
of $\left<y'\right>_{\bar{\nu}}$ on $\bar{\nu}$ that can
be well approximated by
\begin{eqnarray}
\left<y'\right>_{\bar{\nu}} \propto e^{-(1 -
\xi)\bar{\nu}} ,
\label{5.18}
\end{eqnarray}
for $\bar{\nu} \geq 2$.  This gives the fractional
momentum loss per collision
\begin{eqnarray}
{1 \over \left<y'\right>_{\bar{\nu}}} {d \over d\bar{\nu}}
\left<y'\right>_{\bar{\nu}} = - (1 - \xi) = - 0.15.
\label{5.19}
\end{eqnarray}
If we related $\bar{\nu}$ to average nuclear path length
$L$ by $\bar{\nu} = \sigma_{pp}t_A$ and $t_A =
A\rho L$, where $\rho = {4\pi \over 3}R^3_A$, then for
$\sigma_{pp} = 30$mb we have $\bar{\nu} \simeq
0.4 L$ with $L$ in fm.  Thus, if we define the
degradation length  $\Lambda$ by
\begin{eqnarray}
\left<y'\right> = e^{-L/\Lambda},
\label{5.20}
\end{eqnarray}
then
\begin{eqnarray}
\Lambda = [0.4 (1 - \xi)]^{-1} \simeq 17 \, {\rm fm} .
\label{5.21}
\end{eqnarray}
This gives an estimate of how far a proton must travel in a
nuclear medium in order to lose its momentum by a factor
of $e^{-1}$.

The value of $q = 0.37$ for flavor-flip probability may
at first sight appear to be surprisingly large.  However,
if it is regarded as an effective way of accounting for
resonance production, it becomes quite acceptable.  To
see that, we first state that resonance production, which
we have not taken into account explicitly, can easily
produce neutron from $\left.|uud\right>$ through
$\Delta^+
\rightarrow n + \pi^+$.  The process can be depicted
by a dual diagram, as shown in Fig.\ 6(a).  Such a
leakage of $+$ charge through the emission of $\pi^+$
is equivalent to a flavor flip, which changes $UUD$ to
$DUD$, symbolized by a square box in Fig.\ 6(b), and
the favored process, for example, of having valence
quarks changes from $uud$ to $dud$.  Since a
consideration of resonance production would involve
masses, threshold, polarization, decay distribution, and
other complications, our method of using flavor
changes to account for the effect presents considerable
technical economy.  Even though resonance production
can occur only at hadronization in the end charge
leakage can take place at any point where a projectile
valon interacts with the target; hence, the consideration
in Sec.\ IV leading up to Eq.\ (\ref{4.10}) is an effective
way to take such subprocesses into account.

\section{Pion Production}

Having successfully computed the nucleon distributions,
resulting in the determination of the only two free
parameters in the model, we are now able to predict the
pion distributions without any further ambiguities.  The
NA49 data presented in Ref.\ \cite{bc} do not include the
pion spectra.  The E910 data \cite{bc} do have the pion
distributions in the proton fragmentation region; however,
being at $E_{\mbox{lab}} = 12\, GeV$ the energy is too low
to avoid substantial spill-over of quarks and produced
hadrons from target fragmentation into the $x  > 0$
region.  Without the target fragmentation being subtracted,
as is done for $p-\bar{p}$ in the NA49 data, the E910 data
cannot be compared to the predictions from our model.
We present our result below for a future comparison.

The invariant distribution for pion production is
\begin{eqnarray}
{x \over \sigma_{\mbox{in}}} {d \sigma^{\pi} \over dx}
\equiv H_{\pi}(x) = \int {dx_1 \over x_1} {dx_2 \over x_2}
F_{\pi} (x_1, x_2) R_{\pi}(x_1, x_2, x)
\label{6.1}
\end{eqnarray}
where $F_{\pi} (x_1, x_2)$ is the invariant distribution for
finding a quark at $x_1$ and an antiquark at $x_2$, and
$R_{\pi}$ is the corresponding recombination function to
form a pion.  An important aspect about pion production
concerns the role of gluons, a subject that will be discussed
below in connection with $L(z)$.  For now, we consider the
appropriate forms for $F_{\pi}$ and $R_{\pi}$.

We begin with the valon distribution
$G'_{\bar{\nu}}(y'_1, y'_2, y'_3)$, as given by Eq.\
(\ref{3.8}).  As before, we shall omit the subscript
$\bar{\nu}$, and replace it by the valon labels so that
for the two-valon distributions we have
\begin{eqnarray}
G'_{UU} (y'_1, y'_2) = \int^1_0 dy'_3 G'_{UUD} (y'_1,
y'_2, y'_3),
\label{6.2}
\end{eqnarray}
and similarly for $G'_{UD} (y'_1, y'_3)$ by integrating out
$y'_2$.  The single-valon distributions are involved in the 5
subprocesses that contribute to $F_{\pi}(x_1, x_2)$
shown in Fig.\ 7 for  the production of $\pi^+$.  Of
course, because of the flavor changes, the valon labels
are only indicative of the unchanged components.
More precisely, we can express $F_{\pi}$ in the
moment form, as in Eq.\ (\ref{4.35}), for $\pi^+$
\begin{eqnarray}
\tilde{F}_{\pi^+}(n_1, n_2)&=& [\tilde{G}'_{UU}
(n_1, n_2)+ 2 \tilde{G}'_{UD} (n_1, n_2)
]\{\tilde{V}^f_{11}
\tilde{V}^u_{22}  \}_{12}\nonumber\\
&&+ 2 \tilde{G}'_{UD} (n_2, n_1)\{\tilde{V}^u_{12}
\tilde{V}^u_{21}  \}_{12} + 2 \tilde{G}'_U (n_1
+n_2 - 1)\tilde{W}^{fu}_{1, 12}\nonumber\\
&&+ 2 \tilde{G}'_D (n_1
+n_2 - 1)\tilde{W}^{uu}_{3, 12},
\label{6.3}
\end{eqnarray}
and for $\pi^-$
\begin{eqnarray}
\tilde{F}_{\pi^-}(n_1, n_2)&=& \tilde{G}'_{UU} (n_1,
n_2)\tilde{V}^f_{11}
\tilde{V}^u_{22}  \nonumber\\
&&+ 2 \tilde{G}'_{UD} (n_2, n_1) \{\tilde{V}^f_{21}
\tilde{V}^u_{12}  \}_{12} + 2 \tilde{G}'_{UD} (n_1
,n_2)\{\tilde{V}^f_{11}
\tilde{V}^u_{22}\}_{12}\nonumber\\
&&+ 2 \tilde{G}'_U (n_1
+ n_2 - 1)\tilde{W}^{uu}_{1, 12}+ 2 \tilde{G}'_D (n_1
+n_2 - 1)\tilde{W}^{fu}_{3, 12}.
 \label{6.4}
\end{eqnarray}

For the recombination function we have
\begin{eqnarray}
R_{\pi}(x_1, x_2, x) = {x_1x_2 \over x^2}
G^{\pi}_{U\bar{D}} \left({x_1 \over x}, {x_2 \over
x}\right) ,
\label{6.5}
\end{eqnarray}
where $G^{\pi}_{U\bar{D}}(y_1, y_2)$ is the valon
distribution in a pion.  For the latter we adopt the
same form derived in Ref.\ \cite{rch5}
\begin{eqnarray}
G^{\pi}_{U\bar{D}}(y_1, y_2) = \delta(y_1+ y_2 - 1),
\label{6.6}
\end{eqnarray}
which satisfies
\begin{eqnarray}
 \int^1_0 dy_1 \int^{1-y_1}_0 dy_2
G^{\pi}_{U\bar{D}} (y_1, y_2) = \int^1_0 dy_1
\int^{1-y_1}_0 dy_2 (y_1 + y_2)
G^{\pi}_{U\bar{D}}(y_1, y_2) = 1 .
\label{6.7}
\end{eqnarray}
Consequently, the structure function of the pion ${\cal
F}_{\pi}(x)$, that is  related to the single-valon
distribution $G^{\pi}_U(y)$
\begin{eqnarray}
{\cal F}_{\pi}(x) \propto \int^1_x dy G^{\pi}_U(y)
\label{6.8}
\end{eqnarray}
behaves at large $x$ as $(1-x)^1$, in agreement with
the counting rule $(1-x)^{2 r-1}$, where $r$ is the
number of residual quarks ($r = 1$ for pion and $r =
2$ for proton).

Using Eqs.\ (\ref{6.5}) and (\ref{6.6}) in (\ref{6.1})
yields
\begin{eqnarray}
H_{\pi}(x) = {1  \over  x} \int^1_0 dx_1 \int^{1-x_1}_0
dx_2 F_{\pi}(x_1, x_2) \delta(x_1 + x_2 - x) .
\label{6.9}
\end{eqnarray}
By defining
\begin{eqnarray}
H'_{\pi}(x) = x^3 H_{\pi}(x) ,
\label{6.10}
\end{eqnarray}
we then have for the moments [see Eq.\ (\ref{4.22})]
\begin{eqnarray}
\tilde{H}'_{\pi}(n) &=& \int^1_0 dx_1  \int^{1-x_1}_0
dx_2 F_{\pi}(x_1, x_2)(x_1 + x_2)^n\nonumber\\
&=& \sum_{[n_i]} {n!  \over n_1! n_2!}
\tilde{F}_{\pi}(n_1, n_2), \qquad n = n_1 + n_2,
\label{6.11}
\end{eqnarray}
whereupon Eqs.\ (\ref{6.3}) and (\ref{6.4}) can be
used for the production of $\pi^+$ and $\pi^-$,
respectively.  The inverse transform can be done as
before, using Eqs.\ (\ref{5.8}) and (\ref{5.11}).

Having formulated the procedure to calculate the pion
distribution in the projectile fragmentation region, we
must confront one final issue on the role of the gluons.
Although the gluons carry roughly half the momentum
of a proton, no glueball has ever been seen.  They
therefore hadronize by converting to $q \bar{q}$ pairs,
which subsequently form pions.  We take them into
account by enhancing the sea to saturate the
momentum sum rule \cite{dt}.  That is, for the purpose
of pion production we revise the normalization of the
quark distribution $L(z)$ such that $q \bar{q}$ in the
sea carry all the momentum of the incident proton apart
from the momenta of the valence quarks, leaving
nothing for the gluons.  The average momentum
fraction carried by the valence quark in a valon is
\begin{eqnarray}
\left<z\right>_{\mbox{val}} = \int^1_0 dz K_{NS}(z) =
{a
\over  a + b + 1},
\label{6.12}
\end{eqnarray}
where Eq.\ (\ref{2.9}) has been used.  If we denote the
saturated sea distribution by
\begin{eqnarray}
L_1 (z) = \ell_1 (1 - z)^5 ,
\label{6.13}
\end{eqnarray}
where only the constant factor $\ell _1$ has changed
from Eq.\ (\ref{2.10}), then each sea quark carries on
average a momentum fraction of $\ell _1/6$.  That
is to be identified with $(1 -
\left<z\right>_{\mbox{val}} )/2f$, where $f$ is the
number of flavors.  From Eq.\ (\ref{2.13}) we have
$\left<z\right>_{\mbox{val}}  = 0.52$.  Setting $f = 3$,
we get
\begin{eqnarray}
\ell _1 = 0.48 \, .
\label{6.14}
\end{eqnarray}
With this value in $L_1 (z)$, which is then used in Eqs.\
(\ref{4.11}) - (\ref{4.15}) in place of $L(z)$, we obtain
the appropriate $V$ and $W$ functions that should be
used for the calculation of $\tilde{F}_{\pi ^{\pm}}$ in Eqs.\
(\ref{6.3}) - (\ref{6.4}).

Evidently, there are no parameters to adjust for the
calculation of the pion distributions.  The results are
shown in Fig.\ 8 for $\pi^+$ and $\pi^-$ separately at
$\bar{\nu} = 3.1$ and $6.3$.  For ease of comparison
between $\pi^+$ and $\pi^-$ the same curves are
replotted in Fig.\ 9, where the charge and $\bar{\nu}$
dependences are grouped differently.  No data points
are included because none correspond to proton
fragmentation only and at the $\bar{\nu}$ considered,
as discussed at the beginning of this section.
Nevertheless, if one compares our results to the data of
E910 shown in \cite{bc}, there is rough agreement,
both in normalization and in shape.  Generally
speaking, the difference between $\pi^+$ and $\pi^-$
diminishes as $\bar{\nu}$ is increased, though not as
rapidly as in the E910 data at $E_{\mbox{lab}} = 12
GeV$.  We are confident that our predictions will agree
well with the data when they become available, since
the pion distributions have always demonstrated the
reliability of the recombination model \cite{rch5,dh}.

\section{Conclusion}

We have formulated the projectile fragmentation
problem in $pA$ collisions in the valon model.
Despite the non-perturbative aspect of the problem, the
formulation results in a well-defined procedure of
calculating various contributions to the nucleon and
meson distributions in the proton fragmentation
region.  The nuclear target effects, which present the
only unknown in the model, are summarized by two
parameters $k$ and $q$.  They are determined by
fitting $p-\bar{p}$ and $n-\bar{n}$ distributions for
two values of $\bar{\nu}$.  With those parameters
fixed, the predictions for the pion distributions can be
calculated without any other adjustable parameters.

The results of our calculations have shown that the
NA49 data can be well described by the valon model.
The normalization of the nucleon spectra turns out to
be correct without any freedom for adjustment.  The
shapes of the distributions are also acceptably
reproduced.    The inferred value of $k$ that gives a
quantitative measure of momentum degradation can be
translated to a degradation length $\Lambda$
 in the form $e^{-L/\Lambda}$ for the degraded
momentum fraction with $\Lambda \simeq 17$ fm.

While a value for $\Lambda$ can serve as a succinct
numerical summary of the stopping effect of the
nucleus, we admit that it cannot be inferred directly
from the $pA$ collision data without a detailed
analysis in the framework of the valon model.  This
aspect of the problem is worthy of further attention in
the hope that a degradation length can be extracted by
an appropriate model-independent analysis of the data
on nucleon inclusive cross section.

It should be noted that since the valon model does not
make explicit use of Regge exchanges, it is not capable
of predicting the energy dependence.  It is a model that
should be applied only at asymptotic energies where
scaling behavior prevails.  To compensate for that
drawback, it makes possible the $s$-channels approach
to the calculation of the fragment distributions.

There are obvious directions into which this work can
be extended.  One is to incorporate strangeness and
study the distributions of hyperons and kaons.
Another is to generalize from $pA$ to $AA$ collisions.
From the properties of degradation that this work has
revealed, one is better positioned to assess the extent to
which nuclear matter can be compressed in $AA$
collisions.  Furthermore, with some knowledge about
strangeness production in $pA$ collisions, one can
determine for $AA$ collisions how much strangeness
enhancement is normal and how much anomalous.

Although the valon model represents an approach to
multiparticle production that appears to be orthogonal
to most other approaches based on strings, it should be
recognized that the $s$-channel and $t$-channel
approaches are complementary, not contradictory.  One
may be able to identify diagrams in Figs.\ 3 and 4 that
correspond to baryon junction or diquark breaking.
Just as the notion of duality has benefitted the
investigation of high-energy processes many years
ago that found its idealization in the form of Veneziano
amplitude, which has perfect $s$- and $t$-channel
symmetry, so also here in nuclear processes the
exploration of complementary descriptions of
common as well as unusual phenomena can help to
elucidate the underlying dynamics responsible for
them.

\section*{Acknowledgment} We are grateful to
G.\ Veres for a helpful communication. This work
was supported, in part,  by the U.\ S.\ Department of
Energy under Grant No. DE-FG03-96ER40972.

\newpage
\appendix
\section{Appendix}

In Sec.\ III we have considered the problem of
describing $\nu$ collisions with the target nucleons in
terms of the number of collisions that each valon
experiences.  With the {\it i}th valon encountering
$\nu_i$ collisions, the sum $\sum_i \nu_i$ satisfies
the bounds given in Eq.\ (\ref{3.3}).  In this Appendix
we investigate the phenomenological preference for that
sum within that range.

Let us define the integer $\mu$ by
\begin{eqnarray}
\mu = \nu_1 + \nu_2 + \nu_3
\label{A1}
\end{eqnarray}
so that $\mu$ is bounded by
\begin{eqnarray}
\nu \leq \mu \leq 3 \nu .
\label{A2}
\end{eqnarray}
Since at least one valon must interact with the nucleus,
let that valon be $i = 1$.  Let $p$ be the probability
that either one of the other two valons also interacts.
Furthermore, let $B_{\nu}(\mu)$ denote the
probability that out of $\nu$ independent collisions
the target nucleons encounter, $\mu$ valonic collisions
occur.  For $\nu = 1$, we have $B_1(1) = (1-p)^2,
B_1(2) = 2p(1-p)$ and $B_1(3) = p^2$, which count
the probabilities that valons $i = 2$ and $3$ interact in
addition to the $i = 1$ valon.  Generalizing that to
$\nu$ collisions, we have
\begin{eqnarray}
B_{\nu}(\mu) = {2\nu \choose j} p^j (1 - p)^{2\nu -
j}, \qquad j = \mu - \nu  ,
\label{A3}
\end{eqnarray}
which is a binomial distribution of having $\mu - \nu$
valonic collisions by the $i = 2$ and $3$ valons out of a
maximum of $2\nu$ possible such collisions.  We can
define a generating function from $B_{\nu}(\mu)$:
\begin{eqnarray}
S_{\nu}(z) = \sum^{3\nu}_{\mu = \nu} z^{\mu}
B_{\nu}(\mu) = z^{\nu} \sum^{2\nu}_{j = 0}
z^j {2\nu \choose  j} p^j (1 - p)^{2\nu - j}=
z^{\nu}(1 - p +pz)^{2\nu} = [S_1(z)]^{\nu}.
\label{A4}
\end{eqnarray}

The above consideration can now be applied to
Eq.\ (\ref{3.23}) where $\tilde{G}'_{\nu}(n_1, n_2,
n_3)$ is related to $\tilde{G}(n_1, n_2, n_3)$ under
the assumption of Eq.\ (\ref{3.4}), i.\ e., $\nu = \nu_1 +
\nu_2 + \nu_3$.  That assumption is now liberated by
Eq.\ (\ref{A1}) and (\ref{A2}).  The factor
$[\sum^3_{i = 1} d(n_i)/3]^{\nu}$ in Eq.\ (\ref{3.23})
should thus be replaced by $[S_1(z)]^{\nu}$, which
allows $\mu$ to vary between $\nu$ and $3\nu$.  In
the limit $p = 0, S_1(z)$ becomes $z$ and we recover
the earlier result, if we identify
\begin{eqnarray}
z = {1 \over  3} \sum^3_{i = 1} d(n_i).
\label{A5}
\end{eqnarray}
  Using (\ref{A4}) in
Eq.\ (\ref{3.23}), we finally have, with the help of Eq.\
(\ref{3.8})
\begin{eqnarray}
\tilde{G}'_{\bar{\nu}}(n_1, n_2, n_3) &=&
\sum^{\infty}_{\nu = 1} \tilde{G}(n_1, n_2, n_3)
 B_{\nu}(\nu_1 + \nu_2 +
\nu_3)P_{\bar{\nu}}(\nu)\nonumber\\
&&\cdot \left({1\over  3}\right)^{\nu_1 +\nu_2 +
\nu_3}{(\nu_1 + \nu_2 +
\nu_3)!  \over  \nu_1 ! \nu_2 !
\nu_3!} \prod^3_{i=1}d(n_i)^{\nu} .
\label{A6}
\end{eqnarray}
The summation over $\nu_i$ is included in the sum in
Eq.\ (\ref{3.15}) but without the restriction of
(\ref{3.4}).

We have used the
$\tilde{G}'_{\bar{\nu}}$ given in (\ref{A6})
 to calculate $H'(x)$ for $p-\bar{p}$ and $n-\bar{n}$,
as in Secs. IV and V.  The only difference from before is
that we now have one additional parameter $p$ in
(\ref{A2}) to adjust, which controls the number of
valonic collisions above $\mu = \nu$.  Our best fit of
the data, as in Fig.\ 5, yields $p = 0.05$ with the values
of $k$ and $q$ being essentially the same as in Eq.\
(\ref{5.16}).  Since $p$ is so small, any contribution
from $\mu$ different from $\nu$ can be neglected.
Thus it is the data that instruct us on the suitable range
of values of $\nu_i$, namely:  $\nu_1 + \nu_2
+\nu_3$ is dominantly at its lower bound $\nu$.

\newpage

\begin{figure}[htbp]
\centerline{\psfig{figure=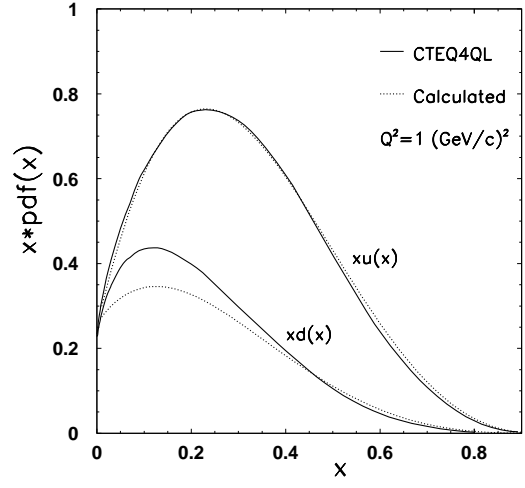,width=0.45\textwidth}}\vspace*{-0.5cm}
\caption{Parton distribution functions at $Q^2 =1
\, ({\rm GeV/c})^2$.  The solid lines are from
CTEQ4QL \cite{cteq}, and the dotted lines are from our
calculation.}
\end{figure}

\begin{figure}[htbp]
\centerline{\psfig{figure=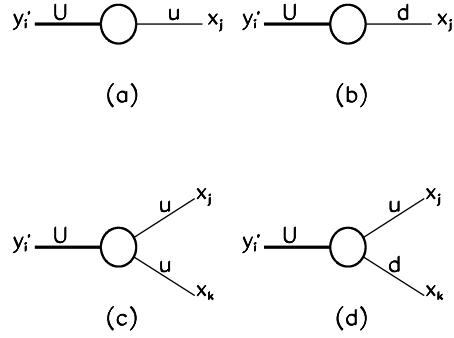,width=0.45\textwidth}}\vspace*{-1cm}
\caption{(a) $V^f_{ij}$, favored quark in a valon,
(b) $V^u_{ij}$, unfavored quark in a valon, (c)
$W^{ff}_{i,jk}$, two favored quarks in a valon, (d)
$W^{fu}_{i,jk}$, a favored and an unfavored quark
in a valon.}
\end{figure}

\begin{figure}[htbp]
\centerline{\psfig{figure=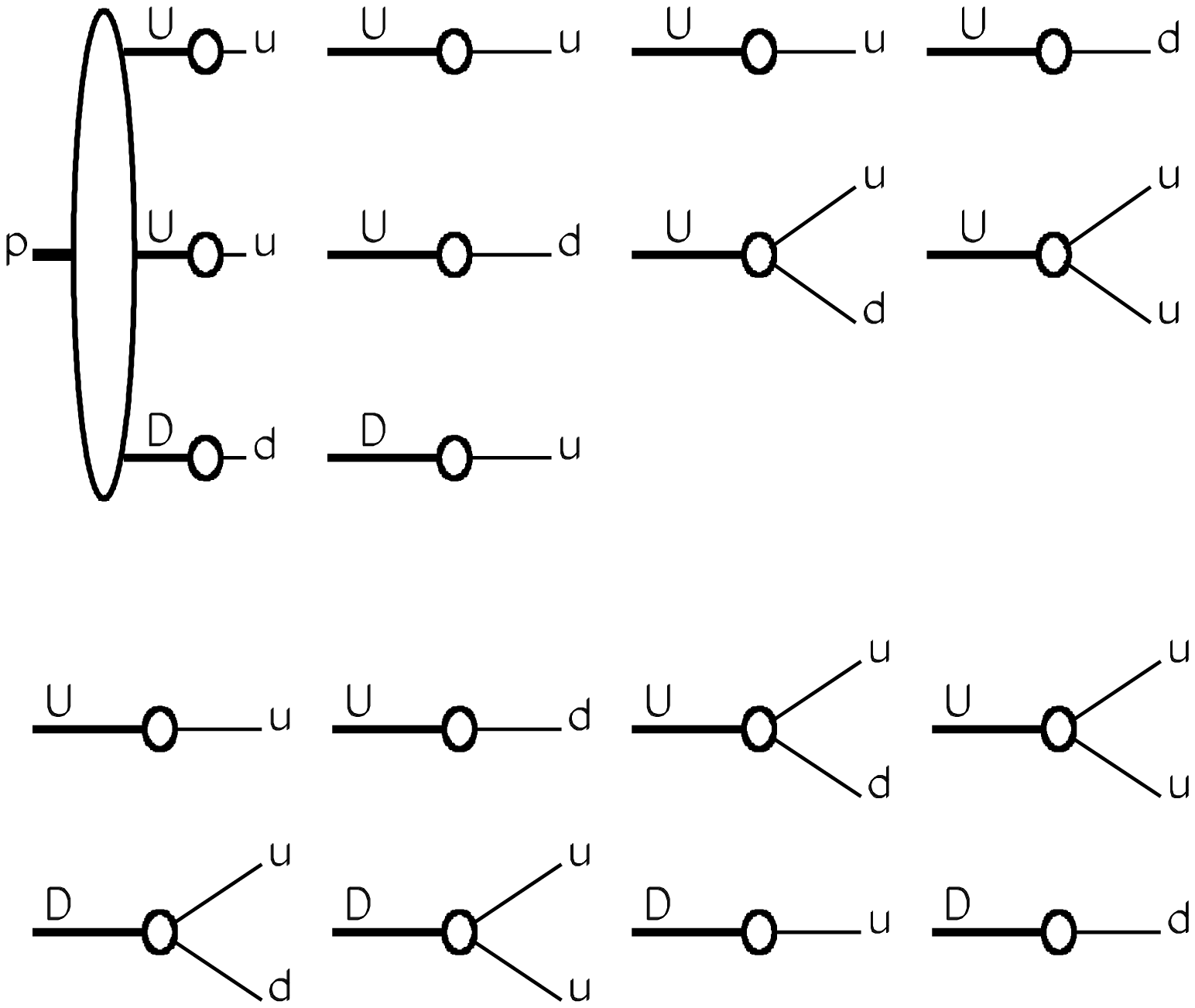,width=0.45\textwidth}}\vspace*{-1cm}
\caption{Eight types of contributions to the quark
 state $uud$ in a proton.}
\end{figure}

\begin{figure}[htbp]
\centerline{\psfig{figure=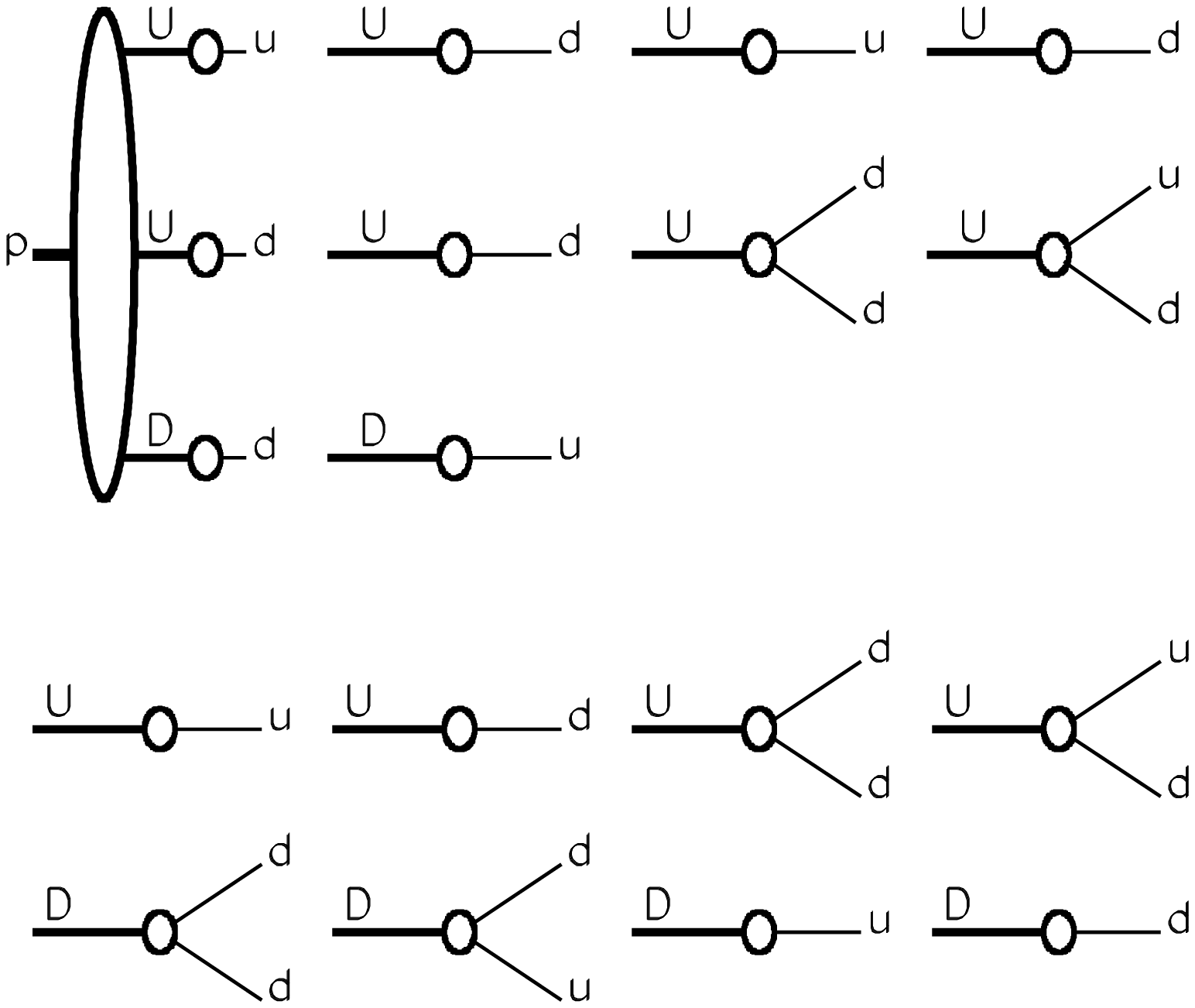,width=0.45\textwidth}}\vspace*{-1cm}
\caption{Eight types of contributions to the quark
 state $udd$ in a proton.}
\end{figure}

\begin{figure}[htbp]
\centerline{\psfig{figure=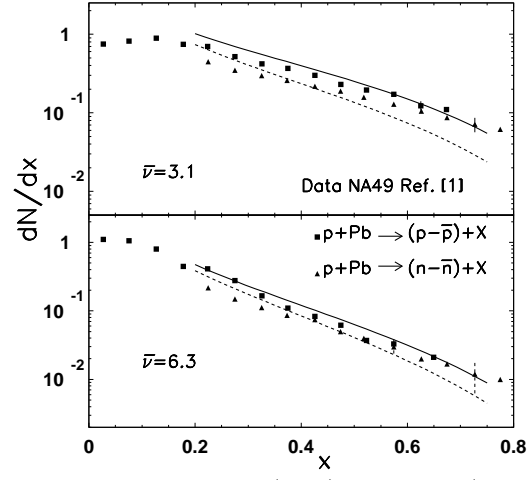,width=0.45\textwidth}}
\vspace*{-1cm}\caption{Inclusive distributions for the production
of $p - \bar{p}$ (solid) and $n - \bar{n}$ (dashed) in
$p$-$Pb$ collisions.  The data are from NA49, reported in
\cite{bc}.}
\end{figure}

\begin{figure}[htbp]
\centerline{\psfig{figure=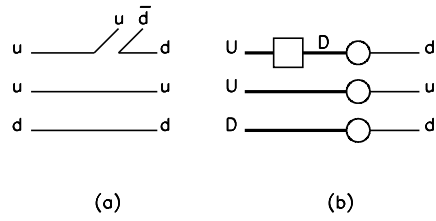,width=0.45\textwidth}}\vspace*{-1cm}
\caption{(a)  A dual diagram that represents the
process $\Delta^+ \rightarrow n + \pi^+$.  (b)  The
square box symbolizes a flavor change from $U$ valon
to $D$ valon before the quark momenta are
determined in the valon model.}
\end{figure}

\begin{figure}[htbp]
\centerline{\psfig{figure=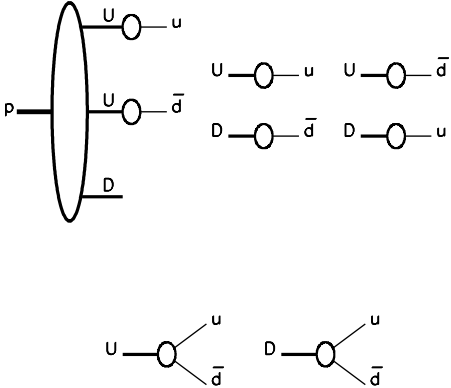,width=0.45\textwidth}}\vspace*{-1cm}
\caption{Five types of contributions to the quark
state $u\bar{d}$ in a proton.}
\end{figure}

\begin{figure}
\centerline{\psfig{figure=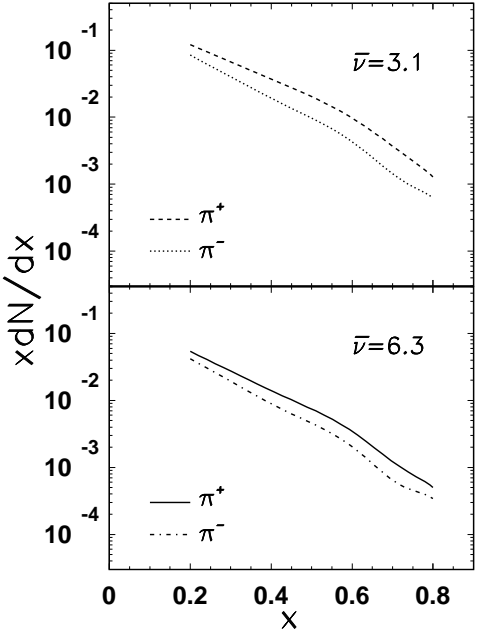,width=0.45\textwidth}}\vspace*{-1cm}
\caption{Inclusive pion distributions of $\pi^+$
and $\pi^-$ at $\bar{\nu}$ = 3.1 and 6.3.}
\end{figure}

\begin{figure}[htbp]
\centerline{\psfig{figure=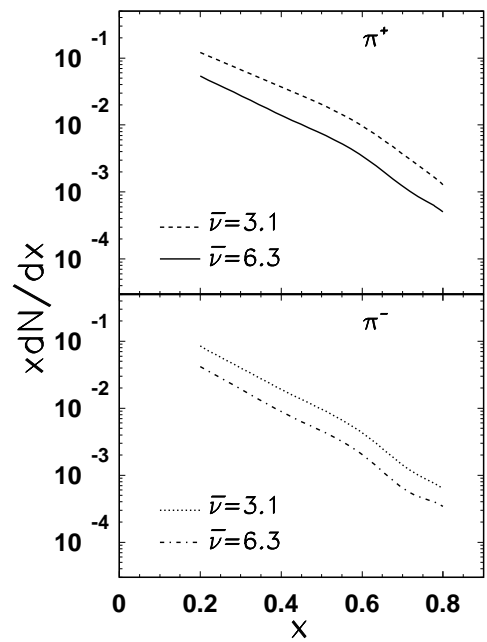,width=0.45\textwidth}}\vspace*{-1cm}
\caption{Same as in Fig.\ 8 but grouped differently.}
\end{figure}

\end{document}